# Muon Colliders, Neutrino Factories, and Results from the MICE Experiment[*]

Daniel M. Kaplan[1, a)]

[1] *Physics Dept., Illinois Institute of Technology, Chicago, Illinois 60616, USA*

[a)]Corresponding author: kaplan@iit.edu

**Abstract.** Muon colliders and neutrino factories are attractive options for future facilities aimed at achieving the highest lepton–antilepton collision energies and precision measurements of parameters of the Higgs boson and the neutrino mixing matrix. The performance and cost of these depend on how well a beam of muons can be cooled. Recent progress in muon cooling design studies and prototype tests nourishes the hope that such facilities can be built starting in the coming decade. The status of the key technologies and their various demonstration experiments is summarized, with emphasis on recent results from the Muon Ionization Cooling Experiment (MICE).

## INTRODUCTION

High-energy lepton colliders have been widely discussed as next-generation facilities to follow up on discoveries at the LHC. Using muons in such machines rather than electrons substantially suppresses radiative processes (which scale with lepton mass as $m^{-4}$), allowing acceleration and collision in rings—greatly reducing both the facility footprint and its construction and operating costs—as well as more-monochromatic collisions and feasibility at much higher energies (10 TeV or more) [1]. Muon decay (mean lifetime = 2.2 μs) complicates beam handling, but also enables stored-muon-beam neutrino factories—the most capable technique for precision measurements of neutrino oscillation [2,3]. Figure 1 compares schematic layouts of these two types of high-energy muon facility; for both, the performance and cost depend on the degree to which a muon beam can be cooled. They have several other common features:

- Both designs start with a high-intensity (megawatt-scale), medium-energy "proton driver" accelerator, illuminating a high-power–capable target (in a heavily shielded enclosure) and copiously producing charged pions, which decay into intense broad-band muon beams.
- First, bunching, then "phase rotation" (reducing the energy spread by accelerating slower muon bunches and decelerating faster ones) prepare the muon beams to be cooled.
- The "initial cooling" stage completes the facility "front ends" [4], which are similar (if not identical) in the two cases.
- Once cooled, the beams are accelerated to the desired energies, injected into storage rings, and circulate for $\mathcal{O}(10^3)$ turns, producing intense and well-characterized neutrino beams, or high-luminosity, high-energy muon–antimuon collisions.

Rubbia has emphasized the importance of muon colliders in the study of the Higgs boson [5]. To test for physics beyond the standard model (SM) requires sub-percent measurements of Higgs branching ratios as well as a precision scan of the resonance line-shape, possible only with *s*-channel production at a 126 GeV muon collider. Studies of the Higgs self-coupling are also needed, requiring a TeV lepton collider.

A natural muon collider staging plan thus emerges [6,7]:

---

[*] To appear in Proceedings, 25th Conference on Application of Accelerators in Research and Industry (CAARI – 2018).

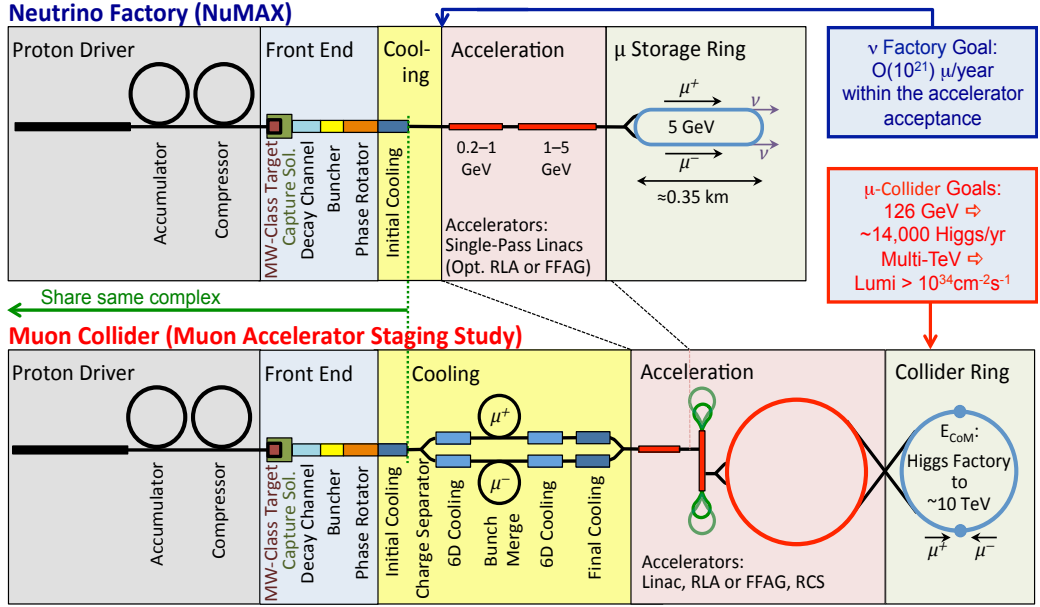

**FIGURE 1.** Comparison of 5 GeV "NuMAX" neutrino factory (top) and muon colliders (bottom). The "front end" (muon production, collection, bunching, phase rotation, and initial cooling) can be very similar for both. It is followed in a neutrino factory by acceleration of the muons to multi-GeV energy and injection into a storage ring, with long straight sections in which muon decay forms intense neutrino beams aimed at near and far detectors. For a muon collider, the front end is followed by six-dimensional cooling, bunch coalescence, and acceleration to high energy (e.g., 63 GeV for a precision "Higgs Factory") for injection into a collider ring, where $\mu^+$ and $\mu^-$ bunches collide for $\sim 10^3$ turns.

1. Start with a neutrino factory,[1] which can do competitive physics with no cooling, and ultimately requires only "initial" cooling by a factor of ~10–50 in six-dimensional (6D) emittance.
2. Then upgrade the facility to a 126 GeV "Higgs Factory" muon–antimuon collider, requiring $\mathcal{O}(10^6)$ or more emittance reduction.
3. Then upgrade to a TeV or multi-TeV collider.

(Also worthy of mention, at low energies, frictional cooling of muons promises smaller and more intense stopping-muon beams [9]. A subject of ongoing R&D at the Paul Scherrer Institute, it may enable enhanced studies of muonium spectroscopy, searches for muonium–antimuonium oscillations, and a unique test of antimatter gravity [10], among other measurements [11].)

## A BRIEF HISTORY OF MUON COLLIDERS

The first discussion of muon colliders in the literature dates from the 1960s [12]. Ionization cooling—the key idea enabling high luminosity—came later [13, 14], and its theory was not fully understood until the 1990s [15].

In the mid-1990s the Muon Collider Collaboration (MCC, including scientists from a number of U.S. national laboratories and universities) formed, producing a report on muon colliders for the 1996 Snowmass meeting [16]. The following year the dedicated neutrino factory was conceived [17], stimulating collaboration expansion and renaming, to the Neutrino Factory and Muon Collider Collaboration [18] (NFMCC), as well as a series of neutrino factory feasibility studies [19–22], the development of the Muon Ionization Cooling Experiment [23], and an annual series of international workshops that continues to this day [24].

In 2006, Fermilab Director Oddone initiated an effort (the Muon Collider Task Force [25], MCTF) to develop a Fermilab-site-specific muon collider proposal, working in collaboration with the NFMCC. In response, the DOE requested a concerted effort, the Muon Accelerator Program (MAP) [26], which began in 2011. MAP concluded in 2017. As of this writing, the publication of the final MAP papers is nearing completion [27], and a new European R&D program on muon colliders is commencing [28].

---

[1] The nuSTORM short-baseline muon storage-ring facility, requiring no new technology, has been discussed as an even earlier step [8].

In its >20 years of activity, the MCC/NFMCC/MCTF/MAP R&D program achieved a number of significant accomplishments, of which the following are some highlights:

1. Design of simple, effective transverse-cooling channels for use in neutrino factories [29] (≈2000–2005);
2. Establishment of neutrino factory feasibility [21] (≈2000–2001);
3. Experimental demonstration of multi-MW liquid-Hg-jet target feasibility [30] (≈2007–2010);
4. Development of successful 6D cooling channel designs [31–34] (≈2005–2015);
5. Experimental demonstration of non-scaling FFAG feasibility [35] (≈2005–2010);
6. Development of successful fast muon acceleration system designs [36];
7. Development of successful collider ring lattices [37];
8. Experimental demonstration of two successful approaches to operation of high-gradient RF cavities in multi-tesla magnetic fields [38,39];
9. Experimental demonstration of transverse-cooling channel feasibility [40].

## IONIZATION COOLING OVERVIEW

While muon colliders and neutrino factories without muon cooling have been considered, at present they appear to fall short in sensitivity compared to designs employing cooling;[2] furthermore, in all studies to date, the cooling channels have been a non-dominant cost driver. Since established methods of beam cooling are ineffective for the muon due to its short lifetime, non-traditional approaches must be used. An approach almost uniquely applicable to muons—ionization cooling [13–15]—works in microseconds, allowing small enough emittances to be reached with $\mathcal{O}(10^{-2\pm1})$ muon survival. In ionization cooling, muons pass through a series of energy absorbers of low atomic number, in a suitable focusing magnetic lattice; the normalized transverse emittance then evolves according to [15]

$$\frac{d\varepsilon_n}{ds} \approx \frac{-1}{\beta^2}\left\langle \frac{dE_\mu}{dx}\right\rangle \frac{\varepsilon_n}{E_\mu} + \frac{\beta_\perp(13.6\,\text{MeV})^2}{2\beta^3 E_\mu m_\mu c^2 X_0} \tag{1}$$

where $\beta c$ is the muon velocity, $\beta_\perp$ the lattice betatron function at the absorber, $dE_\mu/ds$ the energy loss per unit length, $m_\mu$ the muon mass, and $X_0$ the radiation length of the absorber material. (We write $\beta_\perp$ in place of the more usual $\beta_x$ or $\beta_y$ because cylindrically symmetric solenoidal focusing is typically used; thus $\beta_x = \beta_y \equiv \beta_\perp$, and cooling occurs equally in the $x$-$x'$ and $y$-$y'$ phase planes.) To allow iteration, absorbers are interspersed with accelerating cavities.

In Eq. 1, the first term describes cooling, and the second, heating due to multiple Coulomb scattering.[3] The heating term is minimized via small $\beta_\perp$ (strong focusing) and large $X_0$ (low-$Z$ absorber material). For a given cooling-channel design, cooling proceeds towards an equilibrium emittance value at which the heating and cooling terms balance, and beyond which a revised design with lower $\beta_\perp$ is required for continued cooling.

Perhaps counterintuitively, cooling is found to work optimally near ≈ 200 MeV/$c$ momentum [15], where the ionization energy loss rate ("$dE/dx$") in matter is near its *minimum* [42]. This is because of the heating effects of the "straggling tail" at higher momentum and the negative slope of the $dE/dx$ curve below the minimum (which creates problematic, *positive* feedback for energy-loss fluctuations).

While the physics of Eq. 1 is well established, poorly modeled tails of distributions, as well as engineering limitations, could have important impact on ionization cooling-channel cost and performance. For these reasons, an effort was mounted to build and test a realistic section of cooling channel: the international Muon Ionization Cooling Experiment (MICE) [23] (about which more below).

## IONIZATION COOLING STAGES

The evolution of beam emittance in a neutrino factory or muon collider (Fig. 2) dictates that a variety of cooling lattices be employed at the various stages:

---

[2] The LEMMA scheme, in which an ambitious high-intensity positron storage ring with internal target produces cool $\mu^+\mu^-$ pairs just above threshold via $e^+e^-$ annihilation, is currently under study [41]; beam scattering in the target appears to limit its emittance performance, making it perhaps suitable for a TeV muon collider but not for a Higgs Factory.
[3] Analogous to beam cooling by synchrotron radiation, in which energy loss provides cooling, while heating is caused by quantum fluctuations.

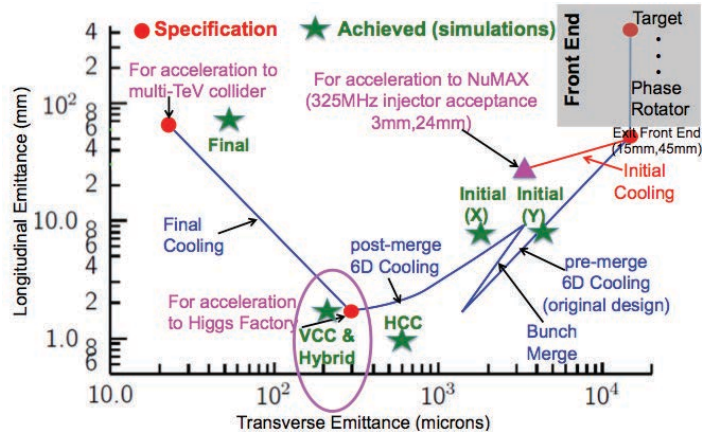

**FIGURE 2.** Cooling "trajectory" in longitudinal and transverse emittance; red points and magenta triangle show Muon Accelerator Program (MAP) emittance goals; green stars, performance achieved in simulation studies.

1. An initial neutrino factory without cooling, producing $\mathcal{O}(10^{20})$ neutrinos/yr, later upgraded to $\mathcal{O}(10^{21})$ neutrinos/yr (in "NuMAX") by adding 6D "initial" cooling [6]. A dual-use linac is employed to accelerate first protons from 3.0 to 6.75 GeV (part of the proton driver), then muons from 1.25 to 5 GeV. Full acceptance by the acceleration system requires muon input transverse and longitudinal emittances of $\approx 3$ and $24\pi$ mm·rad.
2. A Higgs Factory muon collider with exquisite energy spread in support of a precision Higgs-boson line-shape energy scan ($\Gamma_{Higgs,SM} = 4$ MeV). MAP targeted $\approx 0.3/1.5\pi$ mm·mrad transverse/longitudinal emittances, enabling $5 \times 10^{31}$ cm$^{-2}$ s$^{-1}$ luminosity and 5 MeV energy spread, achieved in a series of "6D" cooling channels whose design is introduced below.
3. Above 1 TeV collision energy MAP aimed for transverse/longitudinal emittances of $\approx 0.025/70\pi$ mm·mrad, enabling $10^{34}$ cm$^{-2}$ s$^{-1}$ luminosity. Following the "6D" cooling channels, these parameters were to be achieved by means of "final cooling," which incorporated a significant amount of transverse–longitudinal "reverse" emittance exchange.

Accordingly, there are three main cooling stages in the MAP muon collider approach:
1. Initial Cooling (used also for the neutrino factory)
2. Six-Dimensional (6D) Cooling
3. Final Cooling

## Bunching and Phase Rotation

First, however, to prepare for Initial Cooling, the broad-band beam resulting from pion production and decay is bunched and phase-rotated, by decelerating the faster bunches and accelerating the slower ones [4], in order to bring as many produced muons as possible into the optimal cooling energy range. Taking advantage of the dependence of muon velocity on energy, this can be accomplished using a series of RF cavities of decreasing frequency, ranging from $\approx 500$ down to $\approx 325$ MHz, after first allowing an energy–time correlation to develop within an RF-free drift region (Fig. 3). This process creates a train of approximately monoenergetic alternating $\mu^+$ and $\mu^-$ bunches. The need to minimize muon decay in flight leads to lattice cells that place RF cavities in multi-tesla magnetic fields, thus necessitating the use of normal-conducting RF cavities.

## Initial Cooling

By the year $\approx 2000$, when Neutrino Factory Feasibility Study II (FS2) was carried out [22], successful, purely transverse ionization-cooling lattices had been developed [39]. The FS2 design employed two alternating-solenoid harmonics, allowing small $\beta_\perp$ to be achieved by working between the resulting "$\pi$" and "$\pi/2$" resonances. Figure 4a shows the simplified, more cost-effective design adopted by the International Design Study for the Neutrino Factory (IDS-NF) [3]. In contrast to the bunching and phase-rotation lattice, in such cooling lattices, alternating solenoid-field directions prevent the buildup of a net canonical angular momentum, which would otherwise accrue due to energy loss and re-acceleration within a solenoidal field. (Since solenoids focus in both transverse directions, these lattices are generically referred to as "FOFO," in contrast to FODO alternating-gradient quadrupole lattices.)

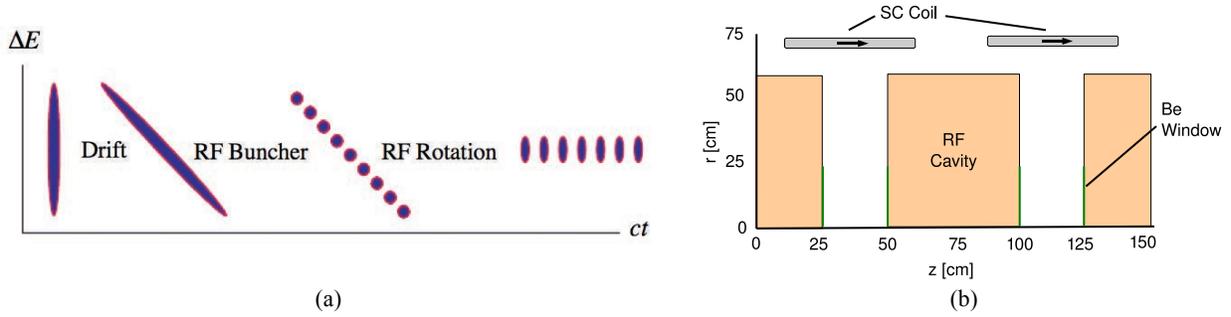

(a)  (b)

**FIGURE 3.** (a) Sketch of buncher and phase-rotation scheme: once an energy–time correlation has developed in a drift region (the higher-energy muons having gotten ahead of the lower-energy ones), a series of RF cavities of decreasing frequencies ranging from ≈ 490 to 326 MHz first bunch the muons, then accelerate the slow ones and decelerate the fast ones, forming an approximately monoenergetic train of alternating $\mu^+$ and $\mu^-$ bunches. (b) Lattice cell used for bunching and phase rotation; note that RF cavity apertures are closed by thin Be windows in order to double the accelerating gradient.

*Emittance Exchange*

Although a transverse cooling channel is simpler than a 6D one, the neutrino factory and muon collider designs can be better unified by employing a 6D (rather than transverse) initial cooling lattice, which (as mentioned) permits cost savings by allowing a dual-use (proton/muon) linac. The inherently transverse ionization-cooling effect can be shared among the transverse and longitudinal phase planes in a lattice in which carefully controlled dispersion causes momentum-dependent path-length through an absorbing medium ("emittance exchange," as shown in Fig. 4b). This results effectively in cooling in all six phase-space dimensions.

At the large transverse emittance of the initial beam, a challenge that must be overcome is to devise a 6D cooling lattice that can simultaneously cool muons of both charge signs, since charge separation of such large-emittance beams would itself be challenging. The challenge is met by the "HFOFO Snake" (Fig. 5) [31], in which small tilts of the solenoids relative to the beam axis, in orientations that rotate about the beam axis by 120° per step, create a small, rotating-dipole field component. This creates periodic orbits and dispersion that are isomorphic, with a half-period offset, for the two muon charges.

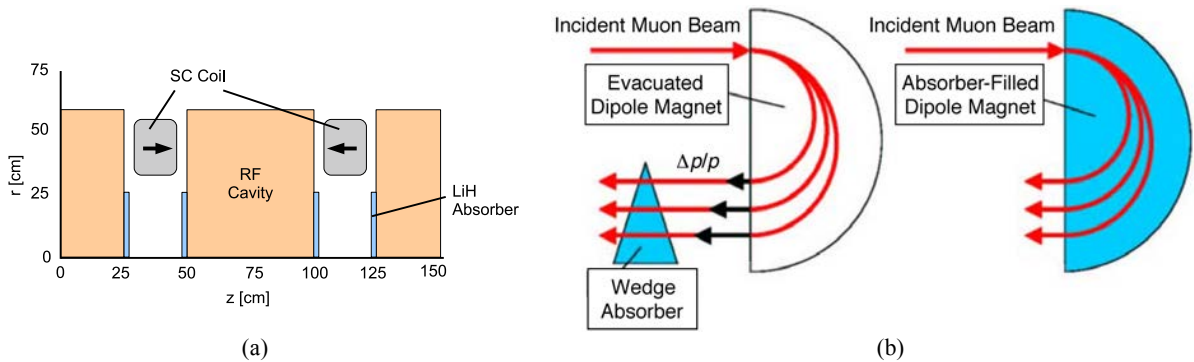

(a)  (b)

**FIGURE 4.** (a) IDS-NF transverse cooling lattice cell, with alternating solenoids to create low-$\beta$ regions between RF cavities and thin, Be-coated LiH absorbers as cavity windows. (b) Two approaches to emittance exchange: in each, an initially small beam with nonzero momentum spread is converted into a more monoenergetic beam with a spread in transverse position. (Figure 4b courtesy of Muons, Inc.)

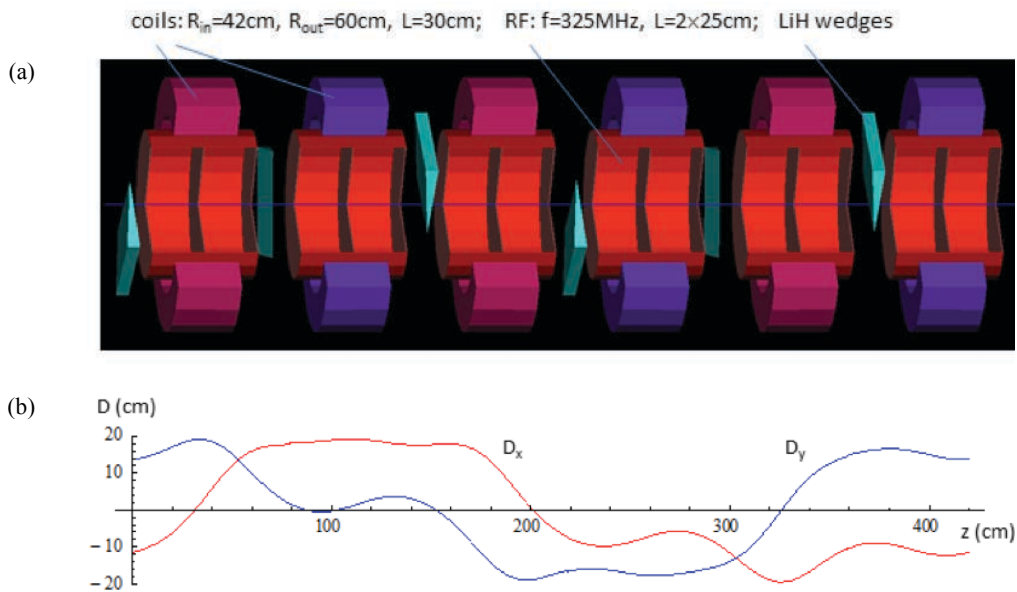

**FIGURE 5.** (a) HFOFO Snake 6D cooling lattice for both charge signs, combining tilted solenoids with LiH wedge and disk absorbers and RF cavities; (b) dispersion vs. position along beam axis.

## 6D Cooling

An overall factor of at least $10^6$ in 6D emittance reduction is necessary for a high-luminosity ($\mathcal{L} \sim 10^{32}$ cm$^{-2}$ s$^{-1}$) Higgs Factory or ($\mathcal{L} \geq 10^{34}$ cm$^{-2}$ s$^{-1}$) TeV muon collider; both transverse and longitudinal cooling are needed. Several approaches towards this goal were developed by the NFMCC [41,42], MCTF, and two small SBIR/STTR-funded R&D firms, Muons, Inc. [43] and Particle-Beam Lasers [44]. This work was continued by many of the same people under the MAP program [27]. Three general approaches were shown to work in simulation: rings, helices, and snakes. Like transverse cooling lattices, most 6D-cooler designs employ superconducting-solenoid focusing and benefit from the ability of such solenoids to accommodate a large aperture, generate low $β$, and focus simultaneously in both $x$ and $y$, enabling compact lattices that minimize muon decay in flight.

Ring coolers were the first to show significant 6D cooling [43], but with insufficient space for beam injection and extraction. This first "in-principle" success led to the development of rings with space allocated for these functions [44,45], and to helices [46,47], with similar beam dynamics as rings, but with open ends for muon ingress and egress, and with reduced absorber and RF-cavity beam loading, since each beam bunch traverses them only once. Helices can also provide faster cooling via focusing strength that increases along the channel, thereby decreasing the equilibrium emittance as the beam is cooled [31]. The Helical Cooling Channel (HCC), based on a Hamiltonian theory [47], employs a combination of "Siberian Snake"-like helical dipole and solenoid fields with a continuous, high-pressure, gaseous-hydrogen absorber, thereby minimizing both the channel length (by eliminating absorber-less RF cavities) and the deleterious effects of windows. Subsequent to the invention of the HCC, it was shown that its required solenoid, helical dipole, and (for increased acceptance) helical quadrupole field components can be produced by a simple sequence of offset current rings [34] (Fig. 6c). As mentioned, the FOFO Snake channel [32] (Figs. 5 and 6b) brings the added benefit of simultaneously accommodating muons of both signs, making it the best choice for Initial Cooling.

While the helical "Guggenheim" design was shown to perform well [31,46], its large volume, together with the need to magnetically shield each turn from its neighbors, led to a search for alternatives. Surprisingly, the same beam dynamics is achievable in a rectilinear geometry, as borne out by detailed simulation studies. Figure 6a shows the lattice geometry, and Fig. 7a a representative performance plot [33]; the 6D cooling factor achieved is $10^5$, with final emittances, 0.28/1.5 mm·rad, exceeding the MAP goal (the lowest point of the Fig. 2 emittance trajectory). (The performance of this design based on vacuum-filled RF cavities has also been demonstrated in a "hybrid" design employing modestly pressurized cavities [48], which may allow higher accelerating gradients.)

# Final Cooling

6D cooling produces a transverse emittance about an order of magnitude too large, and muon bunches shorter than necessary, for a high-luminosity TeV collider, with a 6D emittance an order of magnitude larger than desired. This emittance gap is proposed to be closed via "final cooling" (Fig. 7b), employing reverse emittance exchange, in extremely high-field (30–40 T), small-bore solenoids enclosing $LH_2$ absorbers. Transverse cooling occurs as muon momentum falls towards the Bragg peak of the *dE/dx* curve [49–51], while longitudinal emittance grows due to *dE/dx* positive feedback: on the downslope above the Bragg peak, muons of lower momentum have greater, and higher momentum, smaller, energy-loss rates. Simulation studies of such a channel have come close to the MAP final-emittance goals, falling short by a factor $\approx 2$ in transverse emittance if solenoid fields are limited to 32 T [50,51]. The gap might be fully closed with the use of higher field.[4] Alternatives under study [51] include reverse emittance exchange in wedge absorbers and transverse ionization cooling in quadrupole-focused channels, which can achieve $\beta^* < 1$ cm. Li lenses as combined focusing and absorber elements might also be extendable to these parameters.

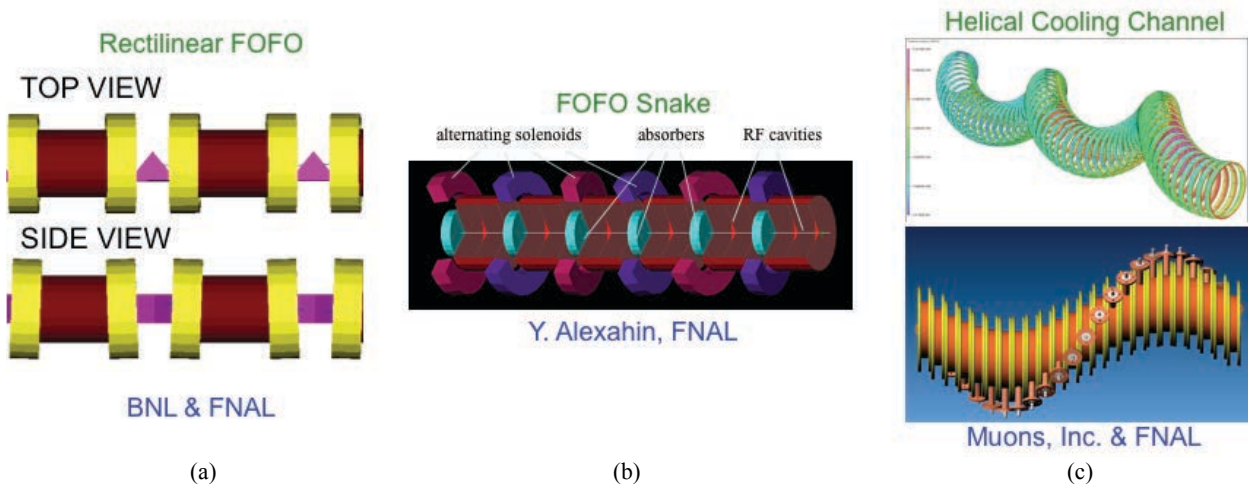

(a) (b) (c)

**FIGURE 6.** Three recent approaches to 6D ionization cooling compared: (a) Rectilinear FOFO, (b) FOFO Snake, and (c) Helical Cooling Channel.

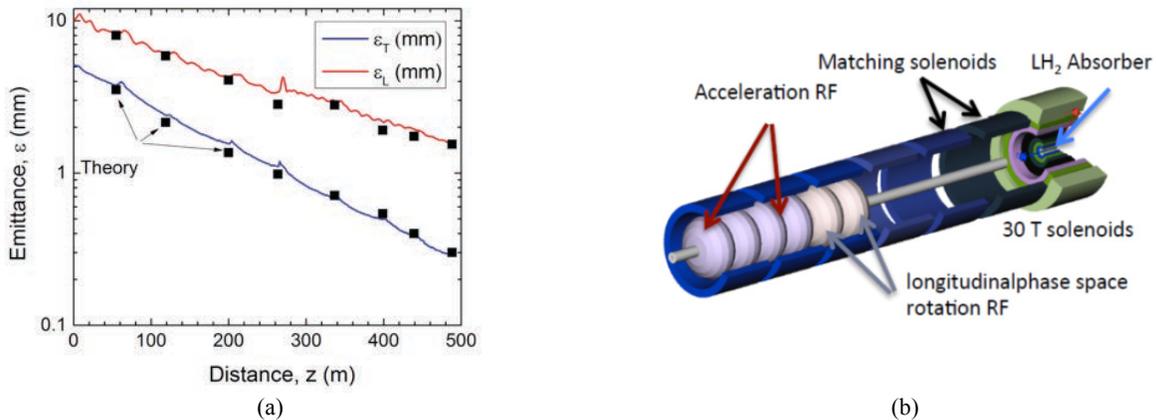

(a) (b)

**FIGURE 7.** (a) 6D cooling performance of the tapered Rectilinear FOFO cooling channel: simulation results (upper and lower curves, respectively) in longitudinal and transverse planes, compared with theoretical predictions (square points). (b) Schematic layout of "final cooling" lattice cell.

---

[4] We note that the National High-Magnetic Field Laboratory now operates a 32 T all-HTS solenoid [52] as well as a 45 T DC hybrid solenoid.

While the above schemes move in the desired direction from the "Higgs Factory" emittance point towards that for a multi-TeV collider, as emphasized by Rubbia [5], higher luminosity at the Higgs mass is desirable, but not at the expense of energy spread. Thus cooling schemes that could lead to lower transverse emittance at the same or lower longitudinal emittance as at the "Higgs Factory" emittance point of Fig. 2 are of interest. An approach that might potentially satisfy Rubbia's requirements is Derbenev's "Parametric-resonance Ionization Cooling" (PIC), in which a resonance (operating in the inverse process to slow extraction) provides smaller $\beta_\perp$ than is practical with magnets [53]; it has been shown to work in principle but still requires a detailed aberration correction scheme to be worked out [54]. Even more speculative ideas have been discussed, including optical stochastic cooling [55] and coherent electron cooling [56] of muons; more work is needed in order to assess their promise.

## MICE

While the principles underlying ionization cooling are well established, ionization cooling channels are tightly packed assemblies that include liquid hydrogen, superconducting magnets, and high-gradient normal-conducting RF cavities in close proximity, with potential safety issues with which little previous experience was available. Moreover, the cooling effect of $\mathcal{O}(10^3)$ ionization cooling cells might be significantly affected by tails of the energy-loss and multiple-scattering distributions. An experimental demonstration was thus deemed essential to further progress.

The Muon Ionization Cooling Experiment (MICE) was proposed as a test of one lattice cell of the Feasibility Study II transverse cooling channel (Fig. 8). Being limited for cost reasons to an $\mathcal{O}(10\%)$ emittance reduction at most, it was conceived as a high-precision measurement of a low-intensity beam, via particle tracking carried out one muon at a time, with the unprecedented emittance resolution of 1‰. It was proposed by an international collaboration [23] and approved in 2003 at Rutherford Appleton Laboratory in the UK. After an extended design, construction, installation, and commissioning process, MICE recorded a substantial dataset (3.5 x $10^8$ events) in 2016–17 with one absorber and no RF cavities. Various absorber materials were studied, including liquid hydrogen and lithium hydride, with a range of beam momentum and lattice focusing strength.

Figures 9 and 10 show some recent MICE results [57]. In Fig. 9 the reconstructed muon amplitude distributions are compared before and after the absorber. Both cooling (density increase at the beam core) and scraping of beam tails are apparent. Due to cost-saving compromises made during MICE construction, the beam transmission through the cooling cell at large amplitude was limited by apertures; furthermore, as is often the case in accelerators, the beam at large amplitude was not well described by a Gaussian. For these reasons, the usual beam-quality figure of merit—total bunch RMS emittance—is not so useful. Thanks to the single-particle measurement capability of MICE, the crucial emittance behavior at and near the core of the beam is nevertheless clearly observable. This can be quantified in various ways, of which two are shown in Fig. 10: subemittance and amplitude ratios. The α-subemittance is defined as the emittance of the central fraction α of the beam; we choose α = 9%, which is the RMS ellipse for a Gaussian beam in the 4 transverse phase-space dimensions ($x$, $p_x$, $y$, $p_y$). Figure 10a compares the behavior of this "$e_9$" subemittance in MICE data and simulation: both display a 10% reduction, highly significant given the small (≈1%) uncertainties. (Note that the small downwards fluctuation of the input beam subemittance relative to that of the simulated beam contributes negligible uncertainty to the measurement.)

Figure 10b shows the behavior of the single-particle amplitude in MICE, for 140 MeV/c beam momentum, with two nominal input-beam emittances and three absorber configurations: (1) absorber empty, (2) 35 cm-thick absorber filled with $LH_2$, (3) 6.5 cm-thick LiH absorber. Shown is the cumulative ratio vs. amplitude of the number of muons downstream to that upstream of the absorber, indicative of the change in phase-space density in the core of the beam. With no absorber, the population of the beam core is negligibly affected. For both the $LH_2$ and LiH absorbers, a clear increase in core density is seen for both input-beam emittances.

## CONCLUSIONS

A largely U.S.-based R&D program explored muon colliders and neutrino factories for over 20 years. Much progress was made, and possible future high-intensity stored-muon facilities now have a much firmer basis than before. Indeed, should a country or region desire to take the next step after LBNF/DUNE, the way to do so is now considerably mapped out; a design project timed for a construction start towards the end of the 2020s seems quite thinkable.

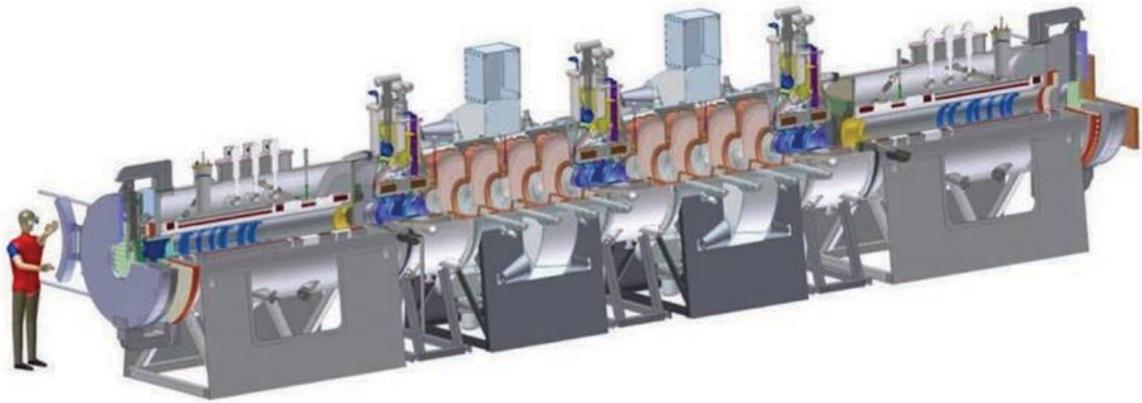

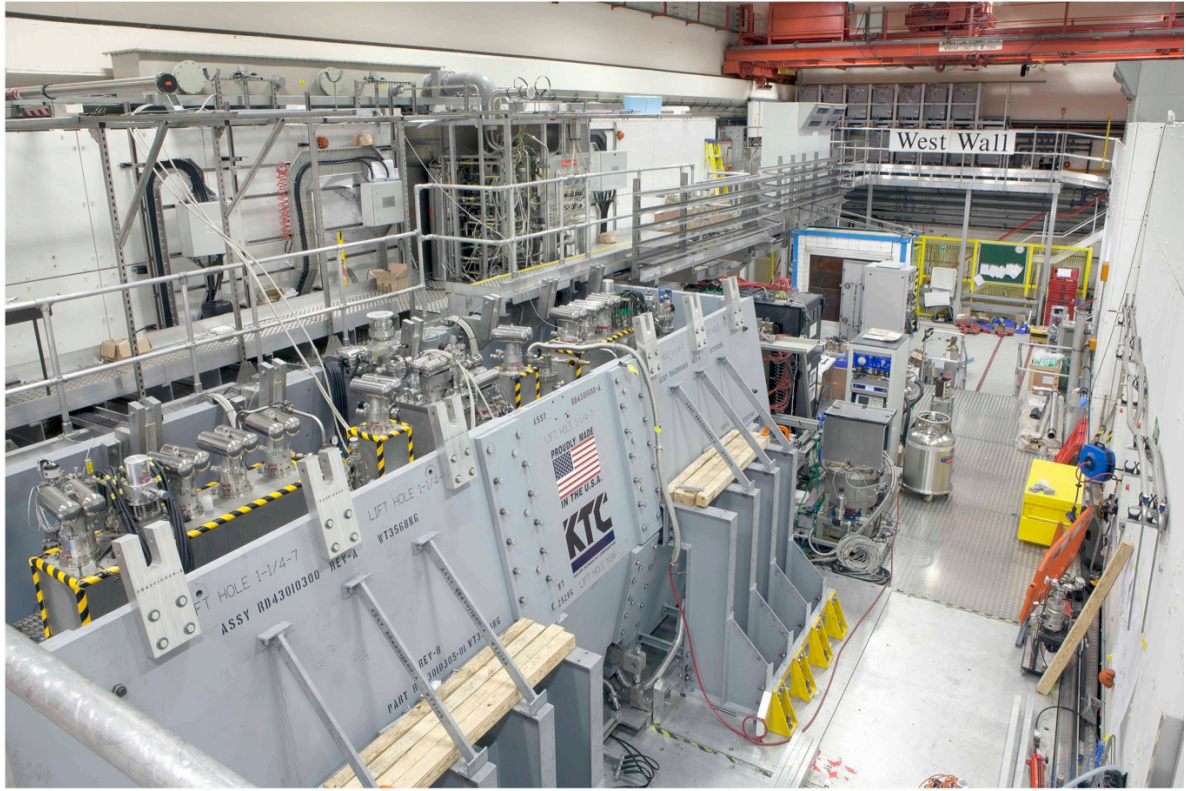

**FIGURE 8.** (a) Cutaway drawing of the MICE experiment as approved, with 3 LH$_2$ absorbers interspersed with two 4-cavity acceleration sections, surrounded by upstream and downstream 4 T solenoidal scintillating-fiber muon-tracking spectrometers, with time-of-flight detectors and calorimeters for particle identification. (b) Photo of MICE as installed, with enclosing iron Partial Return Yoke to protect nearby equipment from solenoid fringe fields.

While the physics case for energy-frontier lepton colliders has not received the boost that discovery of low-mass supersymmetry at the LHC might have given it, it nevertheless looks strong. The upgrade of a neutrino factory into a Higgs Factory and multi-TeV muon collider seems a natural path that the world's high-energy physicists might want to follow in future decades. To be prepared to do so, it would be desirable for a new muon collider R&D program to proceed. Un- (or only partially-) solved problems include whether electron-positron annihilation could become a suitable muon source, how best to implement six-dimensional muon cooling, and how best to attain the desired high luminosities.

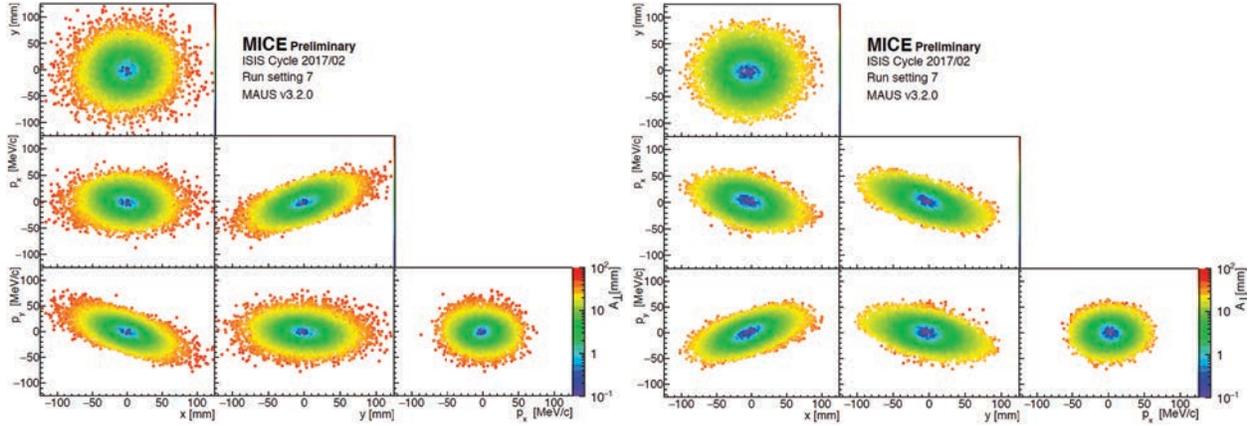

**FIGURE 9.** Beam amplitude for 140 MeV/$c$ beam with 6 mm nominal input emittance vs. phase-space components (a) upstream and (b) downstream of 6.5 cm LiH absorber.

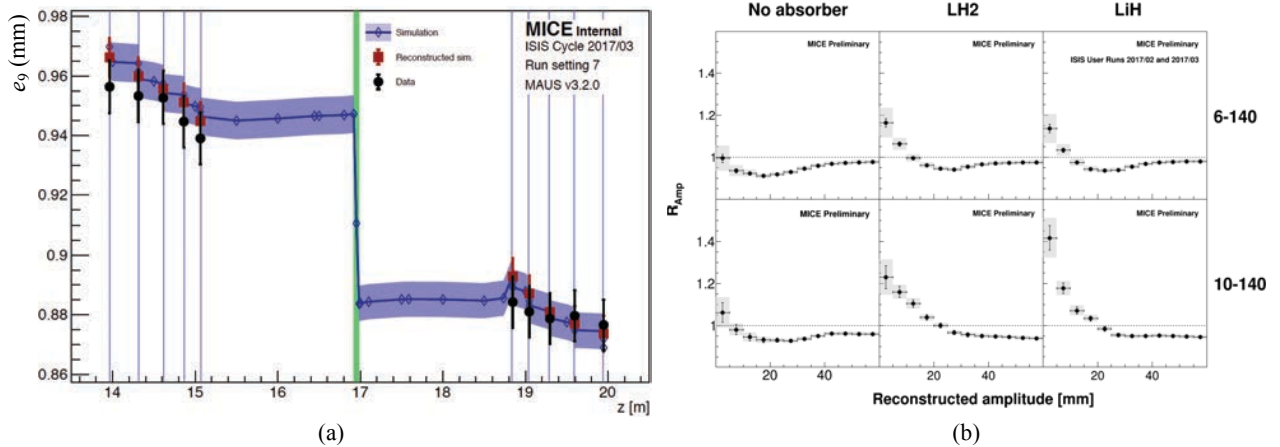

**FIGURE 10.** (a) Behavior of $e_9$ subemittance (emittance of central 9% of beam) in MICE, compared to simulation, for nominal 6 mm input emittance, with 6.5 cm LiH absorber: shaded band shows systematic uncertainty; (b) cumulative amplitude ratios measured in MICE for two nominal input beam emittance and three absorber configurations: solid error bars are statistical uncertainties, gray-shaded bands are systematics.

## ACKNOWLEDGMENTS

The author thanks his colleagues of MAP, the NFMCC, the MICE Collaboration, Muons, Inc., and the MCTF for many stimulating and enlightening conversations and interactions over many years. Work supported by the U.S. Dept. of Energy via the Muon Accelerator Program.